\documentstyle[12pt,aasms4]{article}

\begin{document}

\noindent
{\it Dissertation Summary}

\begin{center}

\title{\large \bf Testing General Relativity in the Strong-Field Regime with Observations of Black Holes in the Electromagnetic Spectrum }

\end{center}

\author{ Tim Johannsen\altaffilmark{1} }

\affil{ Department of Physics and Astronomy, University of Waterloo, \\
200 University Avenue West, Waterloo, ON, N2L 3G1, Canada \\
Perimeter Institute for Theoretical Physics, 31 Caroline Street North, \\
Waterloo, ON, N2L 2Y5, Canada \\
Physics Department, University of Arizona, 1118 East 4th Street, Tucson, AZ 85721, USA
}

\altaffiltext{1}{CITA National Fellow}

\begingroup

\parindent=1cm

\begin{center}

Electronic mail: tjohannsen@perimeterinstitute.ca

Thesis work conducted at: University of Arizona 

Ph.D. Thesis directed by: Dimitrios Psaltis ;  ~Ph.D. Degree awarded: 2012 August 9

{\it Received \underline{\hskip 5cm}}

\end{center}

\endgroup


General relativity, Einstein's theory of gravity, has been tested and confirmed by a variety of different experiments ranging from Eddington's solar eclipse expedition of 1917 to modern observations of double neutron stars (Will 2006, Liv. Rev. Rel., 9, 3). Seldom, however, have these tests probed settings of strong spacetime curvature, and general relativity still stands practically untested in the strong-field regime to date (Psaltis 2008, Liv. Rev. Rel., 11, 9). 

Black holes provide a highly-relativistic environment and a unique laboratory to study some of the most extreme curvature scales in the universe. In this thesis, I create two frameworks for testing general relativity in the strong-field regime with observations of black holes in the electromagnetic spectrum using current or near-future instruments.

In the first part of this thesis, I design tests of the no-hair theorem with observations of black hole accretion flows. According to the no-hair theorem, black holes in general relativity are uniquely characterized in terms of their masses $M$ and spins $a$ and are described by the Kerr metric (e.g., Heusler 1996, Black Hole Uniqueness Theorems, Cambridge Univ. Press). Initially, I investigate a quasi-Kerr metric (Glampedakis \& Babak 2006, Class. Quantum Grav., 23, 4167), which contains an independent quadrupole moment in addition to mass and spin and which is valid for small to moderate values of the spin.

If the no-hair theorem is correct, then any deviation from the Kerr metric has to be zero. If, however, a nonzero deviation is detected, there are two possibilities. Either, the compact object is not a black hole (Collins \& Hughes 2004, Phys. Rev. D, 69, 124022; Hughes 2006, AIP Conf. Proc., 873, 233) or general relativity is only approximately valid in the strong-field regime (e.g., Yunes \& Pretorius 2009, \prd, 79, 084043). In this interpretation, if the object is otherwise known to possess an event horizon, both the no-hair theorem and strong-field general relativity are invalid.

Other parametric deviations from the Kerr metric have been developed (e.g., Manko \& Novikov 1992, Class Quantum Grav., 9, 2477; Collins \& Hughes 2004, Phys. Rev. D, 69, 124022; Yunes \& Pretorius 2009, \prd, 79, 084043; Vigeland \& Hughes 2010, Phys. Rev. D, 81, 024030; Vigeland et al. 2011, Phys. Rev. D., 83, 104027), most of which aim to test general relativity with observations of gravitational waves from extreme mass-ratio inspirals. All of these spacetimes including the quasi-Kerr metric are not ideal for the study of electromagnetic non-Kerr signatures, because they are either valid only for small values of the black hole spin or for small deviations from the Kerr metric or because they contain singularities or closed timelike curves outside of the event horizon. For high values of the black hole spin, however, these pathologies encompass the innermost stable circular orbit hampering the description of accretion flows and the electromagnetic radiation emitted from them.

I, then, construct a Kerr-like black hole spacetime, which is free of such pathologies outside of the event horizon and which is valid for arbitrary values of the spin up to the maximum bound and for large deviations from the Kerr metric (Johannsen \& Psaltis 2011, Phys. Rev. D, 83, 124015). I analyze the properties of this metric as well as the quasi-Kerr metric and show that already moderate changes of the deviation parameters in either metric lead to significant modifications of the observed signals (Johannsen \& Psaltis 2010, ApJ, 716, 187; Johannsen \& Psaltis 2011, Phys. Rev. D, 83, 124015; see, also, Bambi 2011, \prd, 83, 103003; Bambi 2012, \prd, 85, 043002; Bambi \& Barausse 2011, ApJ, 731, 121; Krawczynski 2012, ApJ 754, 133).

First, I apply this framework to the imaging of supermassive black holes using very-long baseline interferometry (VLBI; Johannsen \& Psaltis 2010, ApJ, 718, 446). I show that the shadow of a black hole as well as the shape of a bright and narrow ring surrounding the shadow ("photon ring") depend uniquely on its mass, spin, and inclination (see, e.g., Falcke et al. 2000, ApJ, 528, L13) as well as on the deviation parameter. In addition, I show that the shape of the ring is circular for a Schwarzschild black hole and remains nearly circular for a Kerr black hole within the spin range $|a|\lesssim 0.9M$. For nonzero values of the deviation parameter, however, the ring shape becomes asymmetric, and the degree of asymmetry is a direct measure of the violation of the no-hair theorem (Johannsen \& Psaltis 2010, ApJ, 718, 446).

Sgr~A*, the supermassive black hole in the center of our galaxy, as well as the supermassive black hole in M87 are the prime targets for VLBI imaging observations with the {\em Event Horizon Telescope} (Doeleman et al. 2009, ApJ, 695, 59), a planned global array of (sub-)millimeter telescopes. I argue that the no-hair theorem can be tested with observations of the supermassive black hole Sgr~A* (Johannsen 2012, Adv. Astron., 2012, 1). I estimate the precision of future VLBI arrays consisting of five to six stations and use a Bayesian technique to simulate measurements of the diameter of the ring of Sgr~A* (Johannsen et al., arXiv:1201.0758) with the goal to reduce the correlation of current measurements of the mass and distance of Sgr~A* from the monitoring of stars on orbits around the galactic center (Ghez et al. 2008, ApJ, 689, 1044; Gillessen et al. 2009, ApJ, 692, 1075).

Second, I investigate the potential of quasi-periodic variability observed in both galactic black holes (Remillard \& McClintock 2006, ARAA, 44, 49) and active galactic nuclei (Gierli\'nski et al. 2008, Nature, 455, 369) to test the no-hair theorem in two different scenarios. I consider the diskoseismology model (Wagoner 2008, New Astron. Rev., 51, 828) and a kinematic resonance model (Abramowicz et al. 2003, PASJ, 55, 467) and derive expressions of the dynamical frequencies of the quasi-Kerr and my newly constructed spacetimes. I demonstrate the ability of both models to test the no-hair theorem (Johannsen \& Psaltis 2011, ApJ, 726, 11; Johannsen \& Psaltis, arXiv:1202.6069).

Third, I analyze the prospects of using observations of relativistically broadened iron line profiles to test the no-hair theorem. I simulate iron line profiles over the entire range of spins and demonstrate that deviations from the Kerr metric lead to shifts of the measured flux primarily at high energies as well as in the low-energy tail of the line profiles. I show that these changes can be significant and estimate the required precision of future X-ray missions in order to be able to test the no-hair theorem with fluorescent iron lines for disks of different inclinations (Psaltis \& Johannsen 2012, ApJ, 745, 1; Johannsen \& Psaltis, arXiv:1202.6069).

In the second part of this thesis, I obtain constraints on the size of extra dimensions from the orbital evolution of black-hole X-ray binaries in Randall-Sundrum type braneworld gravity (Randall \& Sundrum 1999, Phys. Rev. Lett., 83, 4690), a string theory-inspired solution of the so-called hierarchy problem. In this model, black holes are unstable (Tanaka 2003, Prog. Theor. Phys. Suppl., 148, 307) and can evaporate into the extra dimension at cosmologically relevant timescales (Emparan et al. 2003, JHEP, 0301, 079).

I derive the resulting rate of change of the orbital period of black hole X-ray binaries in conjunction with magnetic braking and the evolution of the companion star and use existing measurements of the orbital periods of A0620-00 and XTE~J1118+480 to constrain the asymptotic curvature radius of the extra dimension to be less than $161~{\rm \mu m}$ (Johannsen ert al. 2009, ApJ, 691, 997; Johannsen 2009, A\&A, 507, 617). I predicted (at the time of publication) that only one additional measurement of the orbital period of XTE J1118+480 would mark the first detection of orbital evolution in a black hole binary and place the tightest constraint on the size of the extra dimension of $\sim 35~{\rm \mu m}$. Such a measurement was recently obtained confirming the prediction (Gonz\'ales Hern\'andez et al. 2012, ApJ, 744, L25).

Upcoming instruments, such as the {\em Event Horizon Telescope}, {\em Astro-H}, {\em ATHENA}, and the {\em Large Observatory For x-ray Timing} (LOFT) will allow unprecedented views into the workings of black holes across the electromagnetic spectrum. Systematic tests of general relativity in the strong-field regime with electromagnetic observations are within reach.

This work was supported by the NSF CAREER award NSF 0746549 at the University of Arizona and by a CITA National Fellowship at the University of Waterloo. Research at Perimeter Institute is supported by the Government of Canada through Industry Canada and by the Province of Ontario through the Ministry of Research and Innovation.

\end{document}